# Novel Orientational Ordering and Reentrant Metallicity in $K_xC_{60}$ Monolayers for 3 ≤ x ≤ 5

Yayu Wang, R. Yamachika, A. Wachowiak, M. Grobis, K. H. Khoo, D.–H. Lee, Steven G. Louie, and M. F. Crommie

*Department of Physics, University of California at Berkeley, Berkeley, California 94720-7300*
*and Materials Sciences Division, Lawrence Berkeley Laboratory, Berkeley, California 94720-7300*

We have performed local STM studies on potassium-doped $C_{60}$ ($K_xC_{60}$) monolayers over a wide regime of the phase diagram. As K content increases from x = 3 to 5, $K_xC_{60}$ monolayers undergo metal-insulator-metal reentrant phase transitions and exhibit a variety of novel orientational orderings. The most striking new structure has a pinwheel-like 7-molecule unit cell in insulating $K_{4+\delta}C_{60}$. We propose that the driving mechanism for the orientational ordering in $K_xC_{60}$ is the lowering of electron kinetic energy through maximization of the overlap of neighboring molecular orbitals over the entire doping range x = 3→5. In the insulating and metallic phases this gives rise to orbital versions of the superexchange and double-exchange interactions respectively.

Orientational ordering is a classic example of spontaneous symmetry breaking. It occurs in a broad range of materials, such as liquid crystals or quantum magnets, where the basic building block has a rotational degree of freedom. The highly symmetric icosahedral structure of a $C_{60}$ buckyball makes the fullerenes ideal systems for studying novel orientational phases and order-disorder transitions [1]. Orientational order also plays a key role in shaping the rich electronic phase diagram of the doped fullerides that involves superconducting, metallic, insulating, and magnetically ordered phases [2]. Many basic properties of these phases, such as superconductivity and the nature of the metal-insulator transition, are controversial due to the difficulty of disentangling the interactions between orientational order and other degrees of freedom (such as intramolecular vibrations and intermolecular hopping) [3].

Monolayers of K-doped $C_{60}$ provide an excellent two-dimensional laboratory for exploring the complex fulleride physics [4-11]. The importance of the orientational degree of freedom to electronic structure was seen in recent angle-resolved photoemission spectroscopy measurements that reveal significant differences between metallic $K_3C_{60}$ monolayers on Ag(111) and Ag(001), primarily arising from different $C_{60}$ molecular orientations on the two substrates [12]. Long-range orientational ordering in an insulating phase has also been observed in $K_4C_{60}$ monolayers on Au(111) by scanning tunneling microscopy (STM) [13]. This monolayer system was shown to be insulating due to a molecular Jahn-Teller (JT) distortion that is accompanied by cross-like orientational ordering. The local properties of higher-doped $K_xC_{60}$ monolayers with x > 4, however, have not been explored, and the effects of adding electronic charge to the JT-insulating phase and cross-structure are unknown.

In this letter, we report STM studies of the orientational order and electronic structure of $K_xC_{60}$ monolayers in the higher doped regime of $4 \leq x \leq 5$. We find that $K_{4+\delta}C_{60}$ monolayers with $0.1 < \delta < 0.5$ exhibit $C_{60}$ molecules that retain the JT-insulating state but display a complex, seven-sublattice pinwheel-like orientational ordering that has never been observed before. At $x = 5$ the system reenters a metallic phase that exhibits a novel orientational structure strongly related to the $x = 3$ metallic phase. In the JT-insulating $K_xC_{60}$ monolayers a novel superexchange-like mechanism is proposed to be a crucial driving force for the cross and pinwheel phases. This is supported by tight-binding calculations. In the metallic phases the dominant mechanism determining the relative orientation of neighboring molecules is the maximization of the overlap of occupied electron wavefunctions, analogous to the double-exchange interaction.

Our experiments were conducted using a homebuilt ultrahigh vacuum cryogenic STM. $C_{60}$ molecules were deposited onto a clean Au(111) surface from a Knudsen cell evaporator. K atoms were dosed onto the $C_{60}$ monolayer from a K getter. Both K and $C_{60}$ evaporators were pre-calibrated by directly counting the number of atoms/molecules in STM images. The $K_xC_{60}$ monolayer was annealed at $200°C$ for 20 minutes before being cooled to 7 K for STM experiments. After STM measurements at one doping, higher dopings were obtained by adding more K atoms onto the existing monolayer followed by re-annealing. In the highly-doped monolayers with $x > 4.5$, lower annealing temperatures of ~ 140°C are used to avoid K loss. This method enabled us to fabricate $K_xC_{60}$ monolayers with accurate average K contents between $0 \leq x \leq 5$. STM topography was taken in a constant current mode. *dI/dV* spectra were measured through lock-in detection of the ac tunneling current driven by a 450 Hz, 1-10 mV (rms) signal added to the junction bias under open-loop conditions.

Fig. 1(a) shows a typical STM image of the new "pinwheel" phase that occurs in $K_{4+\delta}C_{60}$ (0.1 < δ < 0.5) monolayers. K-deposition in the regime 4 < x < 5 induces a rearrangement of the $C_{60}$ molecules that destroys the long-range cross-phase previously observed in stoichiometric $K_4C_{60}$ (Fig. 1(a) inset), leading to a novel pinwheel-like local structural unit (highlighted by blue circles). *dI/dV* spectroscopy of the pinwheel molecules (Fig. 1(b)) reveals an electronic structure very similar to that seen for the insulating $K_4C_{60}$ cross-phase. This suggests that the pinwheel molecules in $K_{4+\delta}C_{60}$ still retain the energetically favorable $C_{60}^{4-}$ JT-insulating charge state seen in the stoichiometric $K_4C_{60}$ cross-phase [13]. Thus the difference between the cross and pinwheel phases lies purely in the $C_{60}$ lattice structure and intermolecular orientational order.

As shown in Fig. 2(b), each pinwheel consists of six "wheel" $C_{60}$ molecules arranged hexagonally around a central "pin" molecule (each wheel molecule is rotated by approximately 60° relative to its neighbor). The seven molecules sit on a close-packed triangular lattice with inter-molecule distance *a* ~10.8 Å, which is a denser packing (by ~ 5%) than the more open cross-phase. Clustered pinwheels form ordered domains, as shown in Fig. 2(a). Within each domain the seven-sublattice pinwheels arrange themselves into a $\sqrt{7} \times \sqrt{7}$ *R* 40.9° hexagonal superstructure. This suggests that the pinwheels are the structural unit of a new phase rather than merely defects. Both the number and size of the ordered domains grow continuously as the average K content increases from x ~ 4.1 to x ~ 4.5. This is accompanied by the growth of small $C_{60}$-free voids containing only K atoms on Au(111).

The orientation of the center pin molecule reveals another novel aspect of the pinwheel structure. Due to the three-fold rotational symmetry of the underlying triangular lattice formed by $C_{60}$ molecules, the pin molecules are free to choose among three degenerate orientations. Experimentally we observe that the three possible orientations are randomly realized among the pin molecules down to $T = 7$ K, as shown in Fig. 3(a). This type of partial disorder, directly visualized here, is a hallmark of geometric frustration for orientational ordering on a triangular lattice.

When the K content is increased further to x ~ 5, a new stoichiometric, metallic phase emerges, as shown by the ordered domain of bright molecules in Fig. 3(a). $dI/dV$ spectra (Fig. 3(b)) reveals that the gap-feature seen in the JT-insulating phases disappears and a finite density of states exists at $E_F$, indicating a metallic ground state. This provides evidence that this new phase contains $C_{60}$ molecules in a $C_{60}^{5-}$ charge state where the fifth electron occupies an upper JT-split band, as illustrated in Fig. 3(b) inset. The $K_5C_{60}$ molecules form a close-packed triangular lattice with a prominent 2×2 superstructure composed of bright "tri-star" features. Topographic images indicate that each tri-star and its nearest neighbors form a "flower"-like structural unit with the following orientational features: the tri-star (flower-center) exhibits a hexagon facing up, and the six surrounding molecules (flower-petal) exhibit a 6-6 bond facing up (Fig. 3c) [14].

The orientational ordering of $K_5C_{60}$ closely resembles that found in the metallic $K_3C_{60}$ monolayer ( Fig. 3(a) inset) in which $C_{60}$ molecules form a triangular lattice with a $\sqrt{3} \times \sqrt{3}$ superstructure of bright molecules [13]. Fig. 3(c) and (d) present a schematic model of the orientational structure of $K_5C_{60}$ and $K_3C_{60}$. In both phases the $C_{60}$ molecules form a triangular lattice with all neighboring $C_{60}$ molecules (except the central "bright" ones) contacting each other through their equatorial pentagons.

What drives these orientational orderings? The cross and pinwheel patterns are reminiscent of orderings arising from electrostatic quadrupole-quadrupole interactions as proposed by Berlinsky *et al.* [15] (this mechanism has been proposed for other systems as well [16-18]). It is unlikely, however, that electrostatic quadrupole interactions drive the cross and pinwheel phases seen here because (1) the strength of the quadrupole interaction is too weak (estimated to be only ~ 1 meV between nearest $C_{60}$ neighbors), and (2) despite intensive searching via numerical simulation, we cannot reproduce the experimentally observed structures using only a quadrupole-based model [15].

We propose that the main mechanism for all $K_xC_{60}$ orientational phases observed here is the minimization of electron kinetic energy through maximization of the overlap of relevant neighboring molecular orbitals. For the JT-insulating cross and pinwheel-phases ($4 \leq x < 5$) this leads to a novel superexchange-like mechanism, in which virtual hopping of an electron from the HOMO of one $C_{60}^{4-}$ to the LUMO of its nearest neighbor (Fig. 4(b)) gains an energy proportional to $t^2/U$, where $t$ is the hopping amplitude and $U$ is the HOMO-LUMO gap [19]. Here $t$ depends on the overlap between the HOMO and LUMO of adjacent molecules, and thus depends on their relative orientations (Fig. 4(a)).

For the stoichiometric $K_4C_{60}$ insulator, this superexchange favors the cross-phase. Upon additional K deposition to $K_{4+\delta}C_{60}$ the new pinwheel ordering preserves the local $K_4C_{60}$ stoichiometry and benefits energetically from the JT-distortion and superexchange between $C_{60}^{4-}$ molecules by increasing the $C_{60}$ packing density over that of the more open $K_4C_{60}$ cross-phase. This occurs via the segregation of added K atoms into the previously mentioned $C_{60}$-free voids. Adding K atoms to the $K_4C_{60}$ monolayer thus has essentially the same effect as applying an external pressure, a common driver for structural phase transitions.

This interpretation is supported by tight-binding calculations that we have performed to estimate the strength of intermolecular superexchange and its dependence on molecular orientations. $C_{60}^{4-}$ wavefunctions were calculated using the JT-distorted structure obtained previously [13]. For each carbon atom one $\pi$ orbital was kept and only nearest neighbor hopping was allowed. The hopping integral was assumed to be -$2.3 \times (0.145 \text{ nm}/d)^2$ eV, where $d$ is the distance between neighboring carbon atoms [20, 21]. Under this approximation the tight-binding HOMO and LUMO wavefunctions (Fig. 4(a)) are given by linear combinations of $\pi$ orbitals located at different carbon atoms: $|\psi_{HOMO}\rangle = \sum_{i=1}^{60} a_i |\phi_i\rangle$, $|\psi_{LUMO}\rangle = \sum_{j=1}^{60} b_j |\phi_j\rangle$. We have checked that the above parameters give results that agree well with previous DFT-calculations [13].

The strength of the superexchange mechanism is estimated by calculating the HOMO-LUMO hopping integral for neighboring molecules as a function of their relative orientation using $t_{HL} = \langle \psi_{HOMO} | K | \psi_{LUMO} \rangle = \left\langle \sum_{i=1}^{60} a_i \phi_i \middle| K \middle| \sum_{j=1}^{60} b_j \phi_j \right\rangle = \sum_{i,j=1}^{60} a_i^* b_j t_{ij}$, where $t_{ij} = \langle \phi_i | K | \phi_j \rangle$. Following Ref. [23] we take $t_{ij} = V_0 \frac{r_{ij}}{d_0} \exp[-(r_{ij} - d_0)/\lambda] \theta(d_c - r_{ij})$, where $\theta(x)$ is the step function, $r_{ij} = |\vec{r}_i - \vec{r}_j|$, $\lambda = 0.45$ Å, $d_0 = 1.54$ Å, $V_0 = 6.2$ eV, and $d_c = 6.5$ Å [21]. We have computed the energy $-t^2/U$ between two adjacent molecules in the unit cell of the x = 4 cross-structure (calculations for the more complex pinwheel-phase have not been performed). We summed the contributions from the two possible HOMO wavefunctions and used the value $U = 200$ meV (the experimental HOMO-LUMO gap). Both molecules were rotated in a plane and $-t^2/U$ values for different orientations have been plotted in Fig. 4(c). Here $\Delta\phi_1$ and $\Delta\phi_2$ are the angles of the two molecules relative to the observed cross-phase orientation (the cross-phase reference angle is defined as (0, 0)). We find (0,0) to be in the trough of a local minimum in $-t^2/U$, with a value of -15 meV. This represents a significant energy gain for the cross-phase structure, and is much larger than the estimated electrostatic quadrupole interaction. Although there are other local minima in the energy landscape, we take this level of agreement as an encouraging sign for our superexchange proposal.

In the metallic $K_3C_{60}$ and $K_5C_{60}$ phases the superexchange mechanism is not applicable. The likely mechanism for orientational ordering here is enhancement of first-order electron hopping via maximization of the overlap of occupied electron wavefunctions, analogous to the double-exchange interaction. DFT calculations suggest that in the metallic phases the occupied state density peaks near the $C_{60}$ pentagons [14, 19], hence it is energetically favorable to have close intermolecular contact between pentagons to maximize electron hopping. As shown in Figs. 3(c) and (d), the orientational structures of both metallic phases indeed satisfy this requirement. The complex superstructures are typical signatures of geometric frustration on a triangular lattice.

In conclusion, we find that $K_xC_{60}$ monolayers undergo a reentrant metal-insulator-metal transition as x is varied from 3 to 5. Each electronic phase has a novel orientational ordering associated with it. This includes a highly complex, seven-sublattice pinwheel orientational structure in the insulating $K_{4+\delta}C_{60}$ phase. The proposed superexchange and double-exchange mechanisms highlight the close interplay between orientational order and electronic structure in the fullerides.

This work was supported in part by NSF Grant No. EIA-0205641 and by the Director, Office of Energy Research, Office of Basic Energy Science, Division of Material Sciences and Engineering, U.S. Department of Energy under contract No. DE-AC03-76SF0098. Y. W. acknowledges a research fellowship from the Miller institute for Basic Research in Science.

**Figure Captions:**

FIG. 1. (a) STM topograph (V = -0.2 volts, I = 5 pA) of a $K_{4+\delta}C_{60}$ monolayer ($\delta \sim 0.5$) displaying novel seven-sublattice pinwheel structures (highlighted in blue). Inset: topograph of the stoichiometric $K_4C_{60}$ monolayer shows long-range ordered "cross-phase". (b) Spatially averaged *dI/dV* spectroscopy (tip stabilization parameter V = 1 volt and I = 50 pA) for a single pinwheel molecule (red solid curve) is seen to be similar to spectroscopy measured for molecules in the $K_4C_{60}$ cross-phase (black curve).

FIG. 2. (a) STM image of a domain of pinwheels coexisting with the $K_4C_{60}$ cross-phase (V = -0.2 volts, I = 5 pA). The pinwheels form a close-packed $\sqrt{7} \times \sqrt{7}$ superstructure with lattice vectors indicated by red arrows. (b) A close-up image and (c) a schematic structure of a single 7-sublattice pinwheel. The JT-distortion for each $C_{60}$ in the sketch is exaggerated. The observed line node in STM topography is represented here by a red line.

FIG. 3. (a) STM topograph of $K_5C_{60}$ "flower" phase exhibiting 2×2 reconstruction. Inset: topograph of the $K_3C_{60}$ monolayer. (b) Spatially averaged *dI/dV* spectra (tip stabilization parameters V = 1 volt and I = 50 pA) of molecules in $K_5C_{60}$ flower phase show finite density of state at $E_F$. Inset: schematic $K_5C_{60}$ electronic structure. (c) and (d) Model structures for the $K_5C_{60}$ and $K_3C_{60}$ phases. Blue dots represent pentagons on the $C_{60}$ equatorial plane. The orientation of the bright molecules in $K_3C_{60}$ is unknown.

FIG. 4. (a) Local electronic density of states for the HOMO and LUMO of a JT-distorted $C_{60}^{4-}$ molecule calculated by tight-binding. (b) Schematic diagram of the proposed superexchange mechanism between two neighboring $C_{60}^{4-}$ molecules. (c) The orientational dependence of $-t^2/U$ obtained from tight-binding calculations for adjacent molecules. The angle (0, 0) corresponds to the experimentally observed cross-phase.

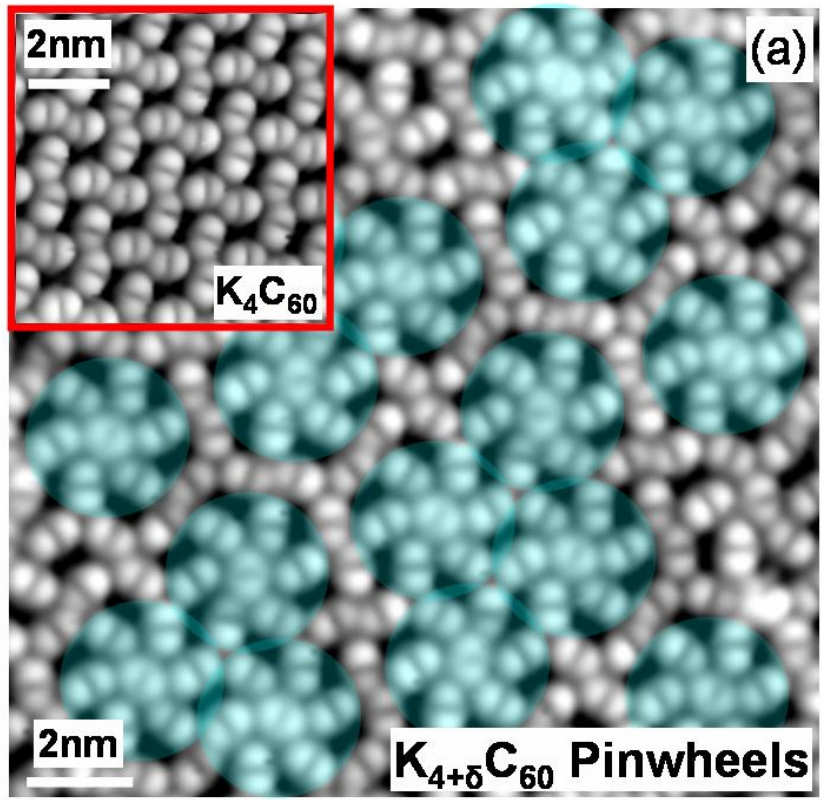

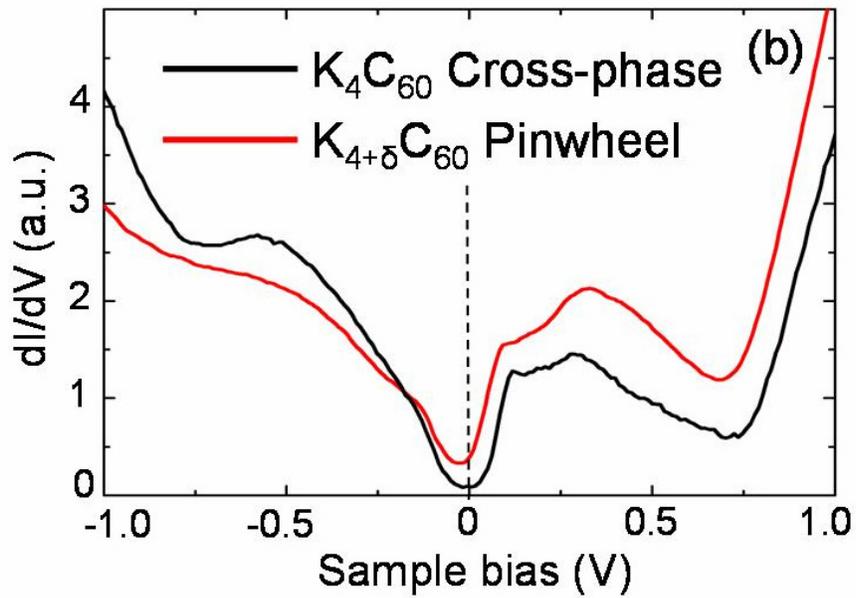

**Figure 1**

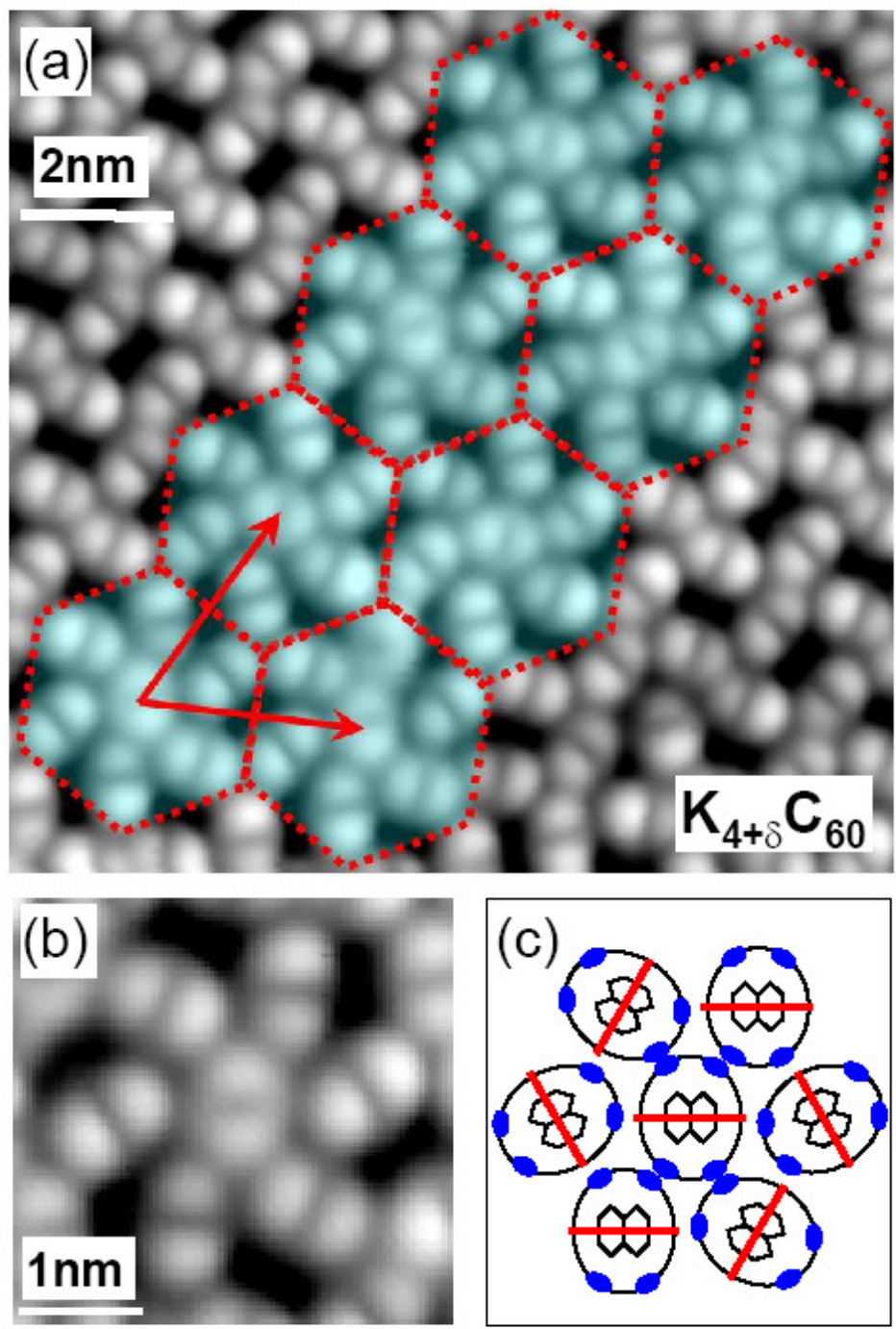

**Figure 2**

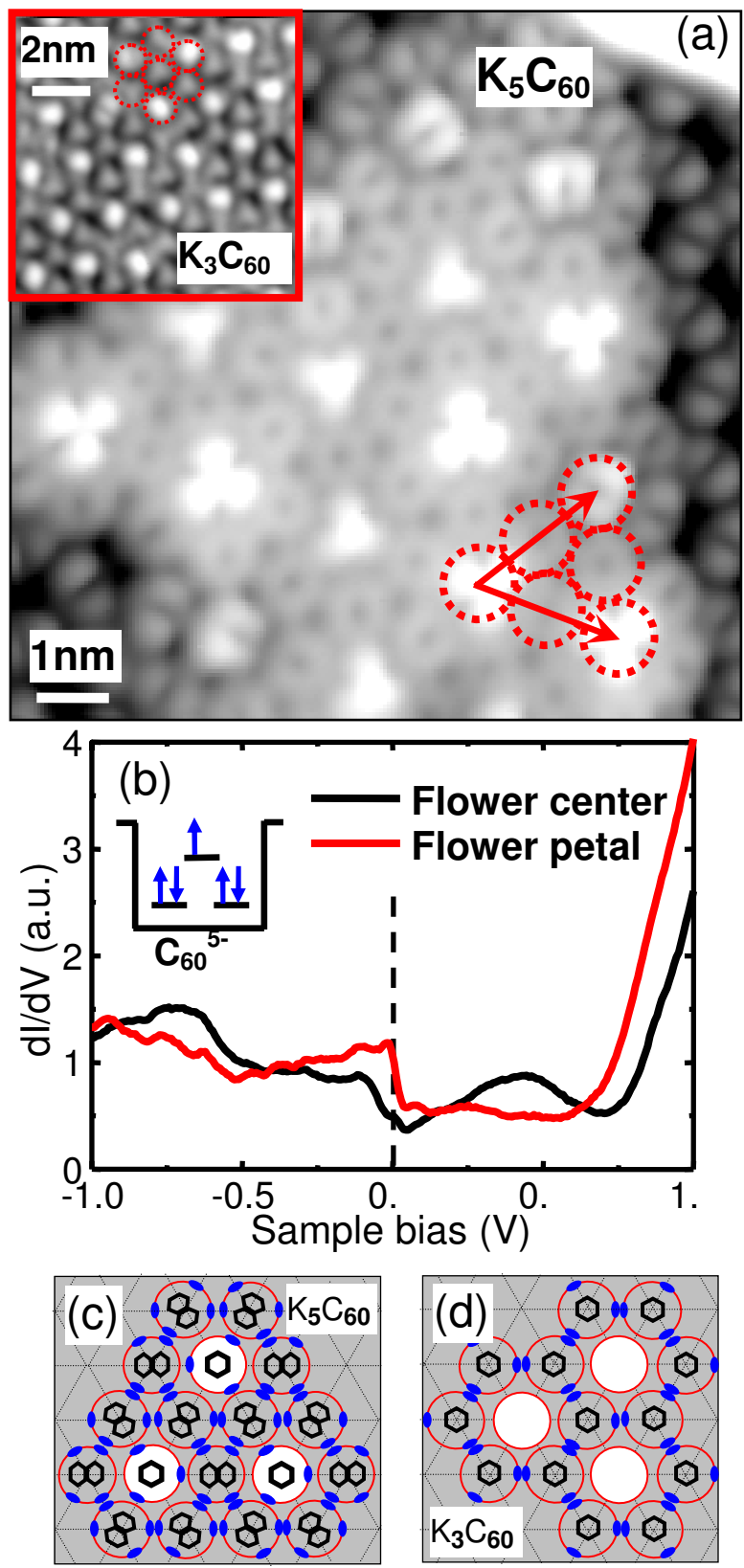

**Figure 3**

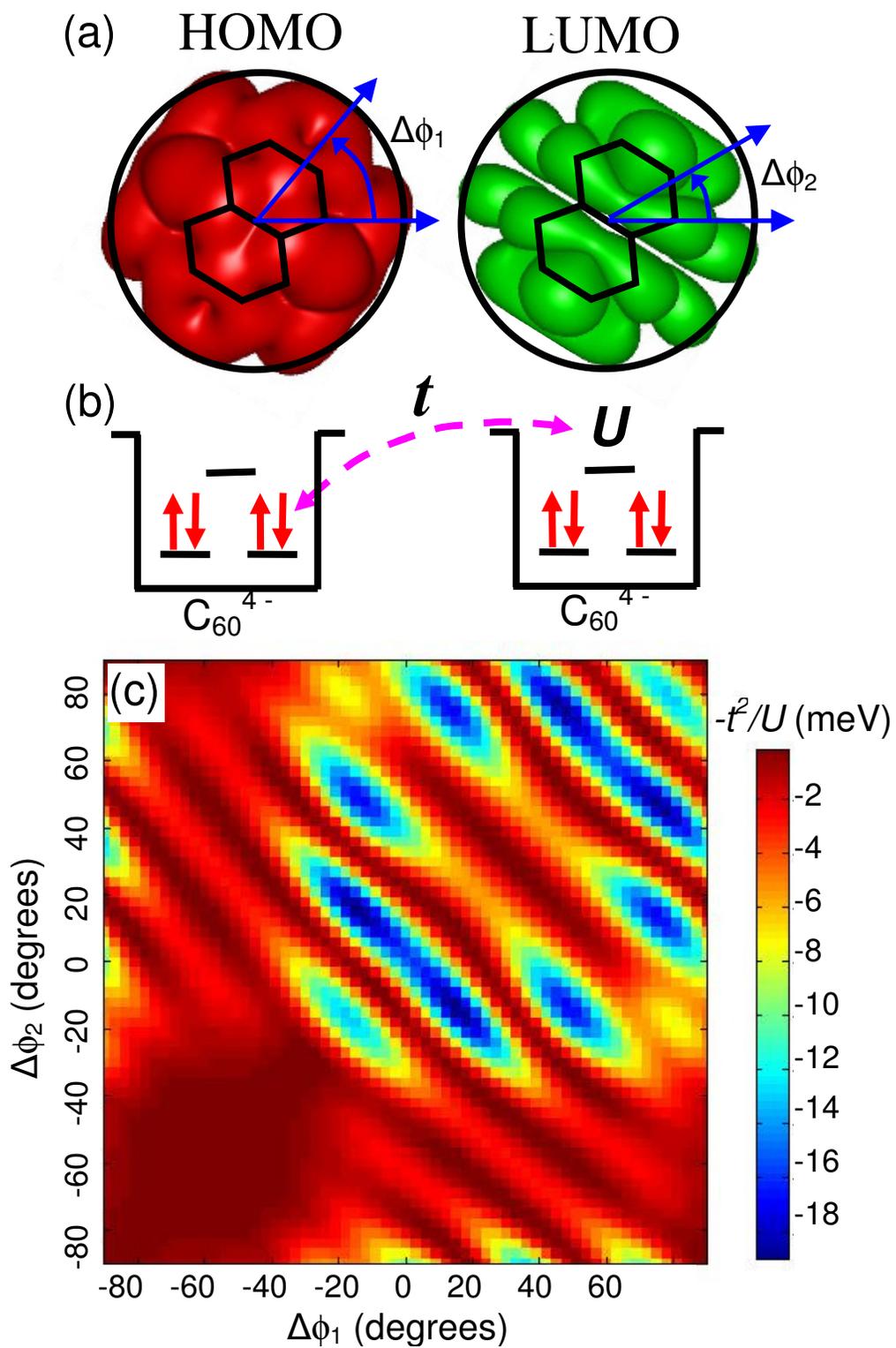

**Figure 4**